\documentclass[12pt]{iopart}

\newcommand{\PRA}{{\it Phys. Rev.} A }

\newcommand{\OC}{{\it Opt. Commun.} }
\newcommand{\JOSAB}{{\it J. Opt. Soc. Am.} B }
\newcommand{\mycomm}[1]{}
\def\COMMENTS{\renewcommand{\mycomm}[1]{\par\noindent\textcolor{red}{\emph{##1}}}}
\newcommand{\ashcomm}[1]{}
\def\COMMENTS{\renewcommand{\ashcomm}[1]{\par\noindent\textcolor{blue}{\emph{##1}}}}

\newcommand{\UQ}{ARC Centre of Excellence for Quantum-Atom Optics, 
School of Physical Sciences, University of Queensland, Brisbane, 
Qld 4072, Australia.}
\usepackage{graphicx}
\usepackage{bm}
\usepackage{iopams}
\usepackage[dvips]{color}

\COMMENTS

\begin{document}

\title{Continuous variable tripartite entanglement from twin nonlinearities}

\author{M.~K. Olsen and A.~S. Bradley}

\address{\UQ}

\begin{abstract}

In this work we analyse and compare the continuous variable tripartite entanglement available from the use of two concurrent or cascaded $\chi^{(2)}$ nonlinearities. We examine both idealised travelling-wave models and more experimentally realistic intracavity models, showing that tripartite entangled outputs are readily producible. These may be a useful resource for such applications as quantum cryptography and teleportation. 
 
\end{abstract}


\submitto{\jpb}

\ead{mko@physics.uq.edu.au}

\maketitle

\section{Introduction}
\label{subsec:intro}

Entanglement is a property which is central to quantum mechanics, with bipartite entanglement being readily producible experimentally. 
There has been some progress in the production of tripartite entangled beams, although the entanglement is often obtained by mixing squeezed vacua with linear optical elements~\cite{Jing,aoki}. Other methods which create the entanglement in the actual nonlinear interaction have been proposed, using both cascaded and concurrent $\chi^{(2)}$ processes~\cite{Guo,ferraro,Nosso}. In this article we investigate the fundamental limits to the achievable tripartite entanglement available from two processes which utilise concurrent and cascaded nonlinearities. We evaluate the continuous variable tripartite entanglement criteria of van Loock and Furusawa~\cite{vanLoock2003} for two different interaction Hamiltonians and then calculate the performance of the corresponding  intracavity systems which contain the interactions described by these Hamiltonians.

\section{Criteria for tripartite entanglement}
\label{sec:criteria}

We will first describe the inequalities which should be violated to demonstrate that a system demonstrates true continuous variable tripartite entanglement. For three modes described by the annihilation operators $\hat{a}_{j}$, where $j=1,2,3$, we define 
quadrature operators for each mode as
\begin{equation}
\hat{X}_j = \hat{a}_j+\hat{a}_j^\dag,\:\:\:
\hat{Y}_j = -i(\hat{a}_j-\hat{a}_j^\dag),
\label{eq:quaddefs}
\end{equation}
so that the Heisenberg uncertainty principle requires $V(\hat{X}_{j})V(\hat{Y}_{j})\geq 1$. 
A set of conditions which are sufficient to demonstrate tripartite entanglement have been derived by van Loock and Furusawa~\cite{vanLoock2003}, without making any assumptions about Gaussian statistics. This is in contrast to the usual conditions used to demonstrate continuous variable bipartite entanglement, which were developed by Duan \etal using the properties of the covariance matrix for Gaussian variables~\cite{Duan}. Using our quadrature definitions, these conditions give a set of inequalities,
\begin{eqnarray}
\eqalign{
V_{12}=V(\hat{X}_1-\hat{X}_2) + V(\hat{Y}_1+\hat{Y}_2+\hat{Y}_3) \geq& 4,\\
V_{13}=V(\hat{X}_1-\hat{X}_3) + V(\hat{Y}_1+\hat{Y}_2+\hat{Y}_3) \geq& 4,\\
V_{23}=V(\hat{X}_2-\hat{X}_3) + V(\hat{Y}_1+\hat{Y}_2+\hat{Y}_3) \geq& 4,}
\label{eq:tripart}
\end{eqnarray}
where $V(A)\equiv\langle A^2\rangle-\langle A\rangle^2$.
If any two of these inequalities are violated, tripartite entanglement is guaranteed.

We note here that, although states which violate these inequalities are sometimes called continuous variable GHZ states, actual GHZ states rely on perfect correlations and are not at all of a statistical nature~\cite{GHZ1,GHZ2}. This terminology would be accurate if the correlations above actually vanished, so that the states involved were eigenstates of amplitude quadrature differences and phase quadrature sums. This would, however, require infinite squeezing of the electromagnetic field which in turn would require infinite energy. Although this situation has been predicted in some works as the result of linearised fluctuation analyses, it always occurs in regions where this form of analysis is not valid. Hence we will not use this terminology in this article. 

\section{Travelling wave models}
\label{sec:quitegood}

In this section we will examine simplified models of two processes which utilise two concurrent or cascaded nonlinearities to produce tripartite entanglement. We will proceed both by solving stochastic equations which are equivalent to the Heisenberg equations of motion derived from the interaction Hamiltonians and also approximate operator equations obtained by ignoring pump depletion. While these procedures are not intended to model realistic physical processes with propagating waves, which would need a more complicated treatment~\cite{fedorente}, they do give some idea of the degree of entanglement available from these Hamiltonians. In the next section we will perform more realistic analyses of the same processes inside pumped optical cavities. 

\subsection{Two cascaded nonlinearities}
\label{subsec:propagate2}

Ferraro \etal have proposed a linked nonlinear process which links five different modes~\cite{ferraro} and was first investigated in terms of photon statistics by Smithers and Lu~\cite{Smithers}. It has recently been analysed in terms of its nonlocal properties by Ferraro and Paris~\cite{frenchconnection}, with the theoretical analyses of all these works using the undepleted pump approximation. However, it is known that the squeezing and bipartite entanglement in travelling-wave $\chi^{(2)}$ processes do not become perfect as the interaction strength increases, but reach some finite limit and then decrease~\cite{revive,turco,shgepr}. We will therefore quantise the pump modes and analyse this system using the full stochastic equations of motion, which must be done numerically. The process involves modes which we will describe using the operators $\hat{b}_{1}(\omega_{1})$, $\hat{b}_{2}(\omega_{2})$, $\hat{a}_{1}(\omega_{3})$, $\hat{a}_{2}(\omega_{4})$ and $\hat{a}_{3}(\omega_{5})$ where the frequencies obey
\begin{eqnarray}
\eqalign{
\omega_{1} = \omega_{3}+\omega_{4},\\
\omega_{5} = \omega_{2}+\omega_{4}.
}
\label{eq:freqferraro}
\end{eqnarray}
The interaction Hamiltonian can then be written
\begin{equation}
H_{\rm int}=\rmi\hbar\left(\chi_{1}\hat{b}_{1}^{\dag}\hat{a}_{1}\hat{a}_{2}+\chi_{2}\hat{b}_{2}^{\dag}\hat{a}_{2}^{\dag}\hat{a}_{3}\right)+{\rm h.c.},
\label{eq:parma}
\end{equation}
which we see describes the same process as ref.~\cite{ferraro} once all the interacting fields are quantised. Note that we have changed the indices of modes 2 and 3 compared with those used by Ferraro \etal so that both our 1 and 2 are produced by the $\chi_{1}$ interaction. This Hamiltonian describes a downconversion process cascaded with a sum-frequency generation process where one of the downconverted modes becomes a pump mode for the frequency generation process.

Before we develop and investigate the full equations from \eref{eq:parma}, it is instructive to examine the analytical solutions which may be obtained using an undepleted pumps approximation. Setting $\kappa_{1}=\chi_{1}\langle\hat{b}_{1}(0)\rangle$ and  $\kappa_{2}=\chi_{2}\langle\hat{b}_{2}(0)\rangle$ as real positive constants, we find the Heisenberg equations of motion,
\begin{eqnarray}
\eqalign{
\frac{\rmd \hat{a}_{1}}{\rmd t} = \kappa_{1}\hat{a}_{2}^{\dag},\\
\frac{\rmd \hat{a}_{2}}{\rmd t} = \kappa_{1}\hat{a}_{1}^{\dag}-\kappa_{2}\hat{a}_{3},\\
\frac{\rmd \hat{a}_{3}}{\rmd t} = \kappa_{2}\hat{a}_{2}.
}
\label{eq:ferraroheisenberg}
\end{eqnarray}
We find that there are two classes of solutions, depending on whether $\kappa_{2}^{2}>\kappa_{1}^{2}$ or $\kappa_{2}^{2}<\kappa_{1}^{2}$. In the first case, where $\kappa_{2}^{2}>\kappa_{1}^{2}$, we set $\Omega=\sqrt{\kappa_{2}^{2}-\kappa_{1}^{2}}$ to find
\begin{eqnarray}
\eqalign{
\hat{a}_{1}(t) = \frac{\kappa_{2}^{2}-\kappa_{1}^{2}\cos\Omega t}{\Omega^{2}}\hat{a}_{1}(0)+\frac{\kappa_{1}\sin\Omega t}{\Omega}\hat{a}_{2}^{\dag}(0)+\frac{\kappa_{1}\kappa_{2}(\cos\Omega t-1)}{\Omega^{2}}\hat{a}_{3}^{\dag}(0),\\
\hat{a}_{2}(t) = \frac{\kappa_{1}\sin\Omega t}{\Omega}\hat{a}_{1}^{\dag}(0)+\hat{a}_{2}(0)\cos\Omega t-\frac{\kappa_{2}\sin\Omega t}{\Omega}\hat{a}_{3}(0),\\
\hat{a}_{3}(t) = \frac{\kappa_{1}\kappa_{2}(1-\cos\Omega t)}{\Omega^{2}}\hat{a}_{1}^{\dag}(0)+\frac{\kappa_{2}\sin\Omega t}{\Omega}\hat{a}_{2}(0)+\frac{\kappa_{2}^{2}\cos\Omega t-\kappa_{1}^{2}}{\Omega^{2}}\hat{a}_{3}(0),
}
\label{eq:ferrarooscillate}
\end{eqnarray}
which, beginning with all the $a$ modes initially as vacuum, gives the solutions for the intensities as
\begin{eqnarray}
\eqalign{
\langle\hat{a}_{2}^{\dag}\hat{a}_{2}\rangle = \frac{\kappa_{1}^{2}\sin^{2}\Omega t}{\Omega^{2}},\\
\langle\hat{a}_{3}^{\dag}\hat{a}_{3}\rangle = \frac{\kappa_{1}^{2}\kappa_{2}^{2}(\cos\Omega t-1)^{2}}{\Omega^{4}},\\
\langle\hat{a}_{1}^{\dag}\hat{a}_{1}\rangle = \langle\hat{a}_{2}^{\dag}\hat{a}_{2}\rangle + \langle\hat{a}_{3}^{\dag}\hat{a}_{3}\rangle.
\label{eq:ansolsintensidadesferraro}
}
\end{eqnarray}
We see that these are the same as the analytical solutions given by Ferraro \etal~\cite{ferraro}, once the change of indices is taken into account.

\begin{figure}
\begin{center} 
\includegraphics[width=0.8\columnwidth]{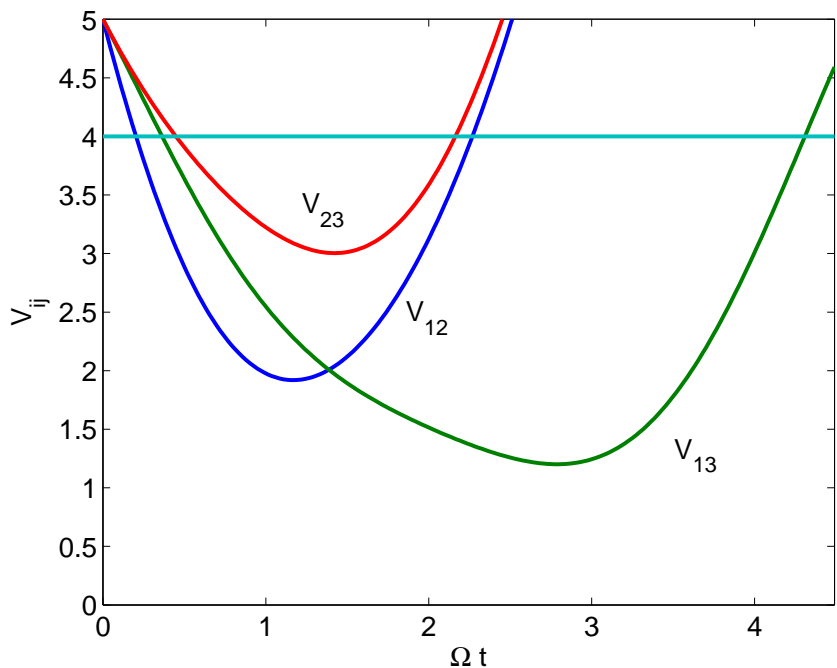}
\end{center} 
\caption{The analytical solutions of the van Loock-Furusawa correlations for the Ferarro scheme, with $\kappa_{2}=1.8\kappa_{1}$. The line at $4$ represents the level beneath which at least two of the correlations must fall to exhibit tripartite entanglement.}
\label{fig:V12anferraro}
\end{figure}

We can use the solutions of \eref{eq:ferrarooscillate} to find expressions for the quadrature variances and covariances,
\begin{eqnarray}
\eqalign{
V(\hat{X}_{1}) = V(\hat{Y}_{1}) = 1+\frac{4\kappa_{1}^{2}\kappa_{2}^{2}(1-\cos\Omega t)-2\kappa_{1}^{4}\sin^{2}\Omega t}{\Omega^{4}},\\
V(\hat{X}_{2}) = V(\hat{Y}_{2}) = 1+\frac{2\kappa_{1}^{2}\sin^{2}\Omega t}{\Omega^{2}},\\
V(\hat{X}_{3}) = V(\hat{Y}_{3}) = 1+\frac{2\kappa_{1}^{2}\kappa_{2}^{2}(1-\cos\Omega t)^{2}}{\Omega^{4}},\\
V(\hat{X}_{1},\hat{X}_{2}) = -V(\hat{Y}_{1},Y_{2}) = \frac{2\kappa_{1}\sin\Omega t}{\Omega^{3}}\left(\kappa_{2}^{2}-\kappa_{1}^{2}\cos\Omega t\right),\\
V(\hat{X}_{1},\hat{X}_{3}) = -V(\hat{Y}_{1},\hat{Y}_{3}) = \frac{2\kappa_{1}\kappa_{2}}{\Omega^{4}}\left[\kappa_{1}^{2}\cos^{2}\Omega t+\kappa_{2}^{2}-(\kappa_{1}^{2}+\kappa_{2}^{2})\cos\Omega t\right],\\
V(\hat{X}_{2},\hat{X}_{3}) = V(\hat{Y}_{2},\hat{Y}_{3}) = \frac{2\kappa_{1}^{2}\kappa_{2}}{\Omega^{3}}\sin\Omega t\left(1-\cos\Omega t\right).
}
\label{eq:ancorrsferraro}
\end{eqnarray}
These expressions contain all the information necessary to express the van Loock-Furusawa correlations, since $V(\hat{X}_{i}-\hat{X}_{j})=V(\hat{X}_{i})+V(\hat{X}_{j})-2V(\hat{X}_{i},\hat{X}_{j})$ and $V(\hat{Y}_{1}+\hat{Y}_{2}+\hat{Y}_{3})=V(\hat{Y}_{1})+V(\hat{Y}_{2})+V(\hat{Y}_{3})+2(V(\hat{Y}_{1},\hat{Y}_{2})+V(\hat{Y}_{1},\hat{Y}_{3})+V(\hat{Y}_{2},\hat{Y}_{3}))$. The expressions for the van Loock-Furusawa correlations obtained by combining these variances and covariances are shown in \fref{fig:V12anferraro}, for $\kappa_{2}=1.8\kappa_{1}$. We see that this Hamiltonian provides tripartite entanglement over a range of scaled interaction time.

\begin{figure}
\begin{center} 
\includegraphics[width=0.8\columnwidth]{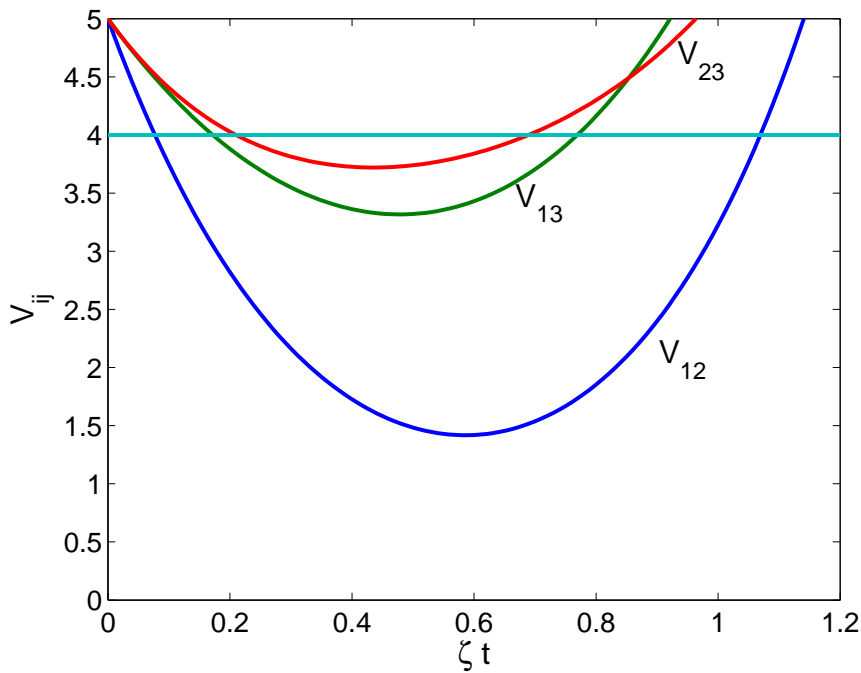}
\end{center} 
\caption{The analytical solutions of the van Loock-Furusawa correlations for the Ferarro scheme, with $\kappa_{1}=1.2\kappa_{2}$. The line at $4$ represents the level beneath which at least two of the correlations must fall to exhibit tripartite entanglement.}
\label{fig:V12hypferraro}
\end{figure}

If $\kappa_{1}^{2}>\kappa_{2}^{2}$ the solutions are not periodic so that the undepleted pumps approximation is of limited validity, being expected to give accurate answers only for short times. However, setting $\zeta=\sqrt{\kappa_{1}^{2}-\kappa_{2}^{2}}$, we find
\begin{eqnarray}
\eqalign{
\hat{a}_{1}(t) = \frac{\kappa_{1}^{2}\cosh\zeta t-\kappa_{2}^{2}}{\zeta^{2}}\hat{a}_{1}(0)+\frac{\kappa_{1}\sinh\zeta t}{\zeta}\hat{a}_{2}^{\dag}(0)+\frac{\kappa_{1}\kappa_{2}(1-\cosh\zeta t)}{\zeta^{2}}\hat{a}_{3}^{\dag}(0),\\
\hat{a}_{2}(t) = \frac{\kappa_{1}\sinh\zeta t}{\zeta}\hat{a}_{1}^{\dag}(0)+\hat{a}_{2}(0)\cosh\zeta t-\frac{\kappa_{2}\sinh\zeta t}{\zeta}\hat{a}_{3}(0),\\
\hat{a}_{3}(t) = \frac{\kappa_{1}\kappa_{2}(\cosh\zeta t-1)}{\zeta^{2}}\hat{a}_{1}^{\dag}(0)+\frac{\kappa_{2}\sinh\zeta t}{\zeta}\hat{a}_{2}(0)+\frac{\kappa_{1}^{2}-\kappa_{2}^{2}\cosh\zeta t}{\zeta^{2}}\hat{a}_{3}(0),
}
\label{eq:ferrarodiverge}
\end{eqnarray}
with the mean intensities being, again with these modes beginning as vacuum,
\begin{eqnarray}
\eqalign{
\langle\hat{a}_{2}^{\dag}\hat{a}_{2}\rangle = \frac{\kappa_{1}^{2}\sinh^{2}\zeta t}{\zeta^{2}},\\
\langle\hat{a}_{3}^{\dag}\hat{a}_{3}\rangle = \frac{\kappa_{1}^{2}\kappa_{2}^{2}(\cosh\zeta t-1)^{2}}{\zeta^{4}},\\
\langle\hat{a}_{1}^{\dag}\hat{a}_{1}\rangle = \langle\hat{a}_{2}^{\dag}\hat{a}_{2}\rangle + \langle\hat{a}_{3}^{\dag}\hat{a}_{3}\rangle.
\label{eq:hypintensidadesferraro}
}
\end{eqnarray}
We note here that these solutions have previously been given by Smithers and Lu~\cite{Smithers} and that we have presented them here because we intend to use them to find analytical expressions for the correlation functions of interest.
In this case the solutions for the variances and covariances are found as
\begin{eqnarray}
\eqalign{
V(\hat{X}_{1}) = V(\hat{Y}_{1}) = 1+\frac{2\kappa_{1}^{2}\left[\kappa_{1}^{2}\sinh^{2}\zeta t +\kappa_{2}^{2}\left(2-2\cosh\zeta t\right)\right]}{\zeta^{4}},\\
V(\hat{X}_{2}) = V(\hat{Y}_{2}) = 1+\frac{2\kappa_{1}^{2}\sinh^{2}\zeta t}{\zeta^{2}},\\
V(\hat{X}_{3}) = V(\hat{Y}_{3}) = 1+\frac{2\kappa_{1}^{2}\kappa_{2}^{2}(\cosh\zeta t-1)^{2}}{\zeta^{4}},\\
V(\hat{X}_{1},\hat{X}_{2}) = -V(\hat{Y}_{1},\hat{Y}_{2}) = \frac{2\kappa_{1}\sinh\zeta t}{\zeta^{3}}\left(\kappa_{1}^{2}\cosh\zeta t-\kappa_{2}^{2}\right),\\
V(\hat{X}_{1},\hat{X}_{3}) = -V(\hat{Y}_{1},\hat{Y}_{3}) = \frac{2\kappa_{1}\kappa_{2}}{\zeta^{4}}\left[(\kappa_{1}^{2}+\kappa_{2}^{2})(1-\cosh\zeta t)+\kappa_{1}^{2}\cosh^{2}\zeta t\right],\\
V(\hat{X}_{2},\hat{X}_{3}) = V(\hat{Y}_{2},\hat{Y}_{3}) = \frac{2\kappa_{1}^{2}\kappa_{2}}{\zeta^{3}}\sinh\zeta t\left(\cosh\zeta t-1\right).
}
\label{eq:hypferrarovars}
\end{eqnarray}
The system also exhibits tripartite entanglement in this regime, as can be seen in \fref{fig:V12hypferraro}. However, as the solutions for the intensities are hyperbolic, they quickly increase to the point where the undepleted pumps approximation will lose its validity.
Neither of the analytic treatments used above is useful for the case where $\kappa_{1}^{2}=\kappa_{2}^{2}$, for which we will employ stochastic integration. 

In developing our full equations of motion, we will follow the approach of Huttner {\em et al.\/}~\cite{Huttner}, (see also Caves and Crouch~\cite{Carlton}) treating the interacting fields in terms of the photon fluxes rather than in terms of energy densities. As stated in Ref.~\cite{Huttner}, this approach avoids problems which could arise, especially with the quantisation volume, if we were to work with the normal Hamiltonian approach. With the appropriate momentum-space operators, we use 
the well-known 
mapping onto stochastic differential equations in 
the positive-P representation~\cite{P+} to calculate the development of the fields as they traverse the medium. We note here that this phase-space representation allows for an exact and complete mapping of our Hamiltonian onto stochastic differential equations. We consider here the case of one dimensional propagation, which is valid for the case of colinear pumping. We also note here that this approach assumes that the medium is not dispersive for the interacting fields, which is a difficult condition to meet with existing materials.  
In this approach, the operator
\begin{equation}
\hat{N}(z_{0},\omega_{m})\equiv \hat{a}^{\dag}(z_{0},\omega_{m})\hat{a}(z_{0},\omega_{m}),
\label{eq:Noperator}
\end{equation}
for example, is the number operator for photons at frequency $\omega_{m}$ which pass through a plane at $z=z_{0}$ during a chosen time interval. The operators $\hat{a}^{\dag}(z,\omega_{m})$ and $\hat{a}(z,\omega_{m})$ then obey bosonic spatial commutation relations,
\begin{equation}
[\hat{a}(z,\omega_{i}),\hat{a}^{\dag}(z',\omega_{j})]=\delta_{ij}\delta(z-z'),
\label{eq:spacecommute}
\end{equation} 
and similarly for the $\hat{b}_{j}$ operators. 
The nonlinear momentum operator for this system is found as
\begin{equation}
\hat{G}_{nl}(z) = \rmi\hbar\left(\chi_{1}\hat{b}_{1}^{\dag}\hat{a}_{1}\hat{a}_{2}+\chi_{2}\hat{b}_{2}^{\dag}\hat{a}_{2}^{\dag}\hat{a}_{3}\right)+{\rm h.c.}
\label{eq:Gferraro}
\end{equation}
As shown by Shen~\cite{Shen}, we can write an equation of motion for the density matrix of the system,
\begin{equation}
\rmi\hbar\frac{\partial\rho(z)}{\partial z}=\left[\rho(z),\hat{G}_{nl}(z)\right],
\label{eq:propagator}
\end{equation}
which allows for the calculation of steady-state propagation, exactly as required for continuous pumping. Physically, the density matrix, $\rho(z)$, describes an ensemble of steady-state systems which has all the statistical properties of the fields at point $z$. \Eref{eq:propagator} provides a full description of the interacting fields of our model, but is extremely 
difficult to solve directly.

Therefore, following the standard procedures~\cite{Crispin}, we map the master equation onto the Fokker-Planck equation for the positive-P pseudoprobability distribution,
\begin{eqnarray}
\eqalign{
\frac{\rmd P}{\rmd z} =& \left\{-\left[\frac{\partial}{\partial\alpha_{1}}\chi_{1}\alpha_{2}^{+}\beta_{1}+\frac{\partial}{\partial\alpha_{1}^{+}}\chi_{1}\alpha_{2}\beta_{1}^{+}
\right.\right.\\
&\left.\left.
+\frac{\partial}{\partial\alpha_{2}}\left(\chi_{1}\alpha_{1}^{+}\beta_{1}-\chi_{2}\alpha_{3}\beta_{2}^{+}\right)+\frac{\partial}{\partial\alpha_{2}^{+}}
\left(\chi_{1}\alpha_{1}\beta_{1}^{+}-\chi_{2}\alpha_{3}^{+}\beta_{2}\right)\right.\right.\\
&\left.\left.
+\frac{\partial}{\partial\alpha_{3}}\chi_{2}\alpha_{2}\beta_{2}+\frac{\partial}{\partial\alpha_{3}^{+}}\chi_{2}\alpha_{2}^{+}\beta_{2}^{+}\right.\right.\\
&\left.\left.
+\frac{\partial}{\partial\beta_{1}}\left(-\chi_{1}\alpha_{1}\alpha_{2}\right)+\frac{\partial}{\partial\beta_{1}^{+}}\left(-\chi_{1}\alpha_{1}^{+}
\alpha_{2}^{+}\right)\right.\right.\\
&\left.\left.
+\frac{\partial}{\partial\beta_{2}}\left(-\chi_{2}\alpha_{2}\alpha_{3}^{+}\right)+\frac{\partial}{\partial\beta_{2}^{+}}\left(-\chi_{2}\alpha_{2}^{+}
\alpha_{3}\right)\right]\right.\\
&\left.
+\frac{1}{2}\left[\frac{\partial^{2}}{\partial\alpha_{1}\partial\alpha_{2}}2\chi_{1}\beta_{1}+\frac{\partial^{2}}{\partial\alpha_{1}^{+}
\partial\alpha_{2}^{+}}2\chi_{1}\beta_{1}^{+}\right.\right.\\
&\left.\left.
-\frac{\partial^{2}}{\partial\alpha_{2}\partial\beta_{2}}2\chi_{2}\alpha_{3}-\frac{\partial^{2}}{\partial\alpha_{2}^{+}\partial\beta_{2}^{+}}
2\chi_{2}\alpha_{3}^{+}\right]\right\}P(\tilde{\alpha},z),
\label{eq:FPmilano}}
\end{eqnarray}
where $\tilde{\alpha}=(\alpha_{1},\alpha_{1}^{+},\alpha_{2},\alpha_{2}^{+},\alpha_{3},\alpha_{3}^{+},\beta_{1},\beta_{1}^{+},\beta_{2},\beta_{2}^{+})$.
As always with the positive-P representation, stochastic averages of products of the variables represent normally ordered operator expectation values, with there being correspondences between $\alpha_{j},\alpha_{j}^{+},\beta_{j},\beta_{j}^{+}$ and $\hat{a}_{j},\hat{a}_{j}^{\dag},\hat{b}_{j},\hat{b}_{j}^{\dag}$.
We now map this Fokker-Planck equation onto the following set of stochastic differential equations in It\^o calculus,
\begin{eqnarray}
\eqalign{
\frac{\rmd\alpha_{1}}{\rmd z} = \chi_{1}\alpha_{2}^{+}\beta_{1}+\sqrt{\frac{\chi_{1}\beta_{1}}{2}}\left(\eta_{1}+i\eta_{3}\right),\\
\frac{\rmd\alpha_{1}^{+}}{\rmd z} = \chi_{1}\alpha_{2}\beta_{1}^{+}+\sqrt{\frac{\chi_{1}\beta_{1}^{+}}{2}}\left(\eta_{2}+i\eta_{4}\right),\\
\frac{\rmd\alpha_{2}}{\rmd z} = \chi_{1}\alpha_{1}^{+}\beta_{1}-\chi_{2}\alpha_{3}\beta_{2}^{+}+\sqrt{\frac{\chi_{1}\beta_{1}}{2}}\left(\eta_{1}-i\eta_{3}\right)-
\sqrt{\frac{\chi_{2}\alpha_{3}}{2}}\left(\eta_{7}-i\eta_{5}\right),\\
\frac{\rmd\alpha_{2}^{+}}{\rmd z} = \chi_{1}\alpha_{1}\beta_{1}^{+}-\chi_{2}\alpha_{3}^{+}\beta_{2}+\sqrt{\frac{\chi_{1}\beta_{1}^{+}}{2}}\left(\eta_{2}-i\eta_{4}\right)-
\sqrt{\frac{\chi_{2}\alpha_{3}^{+}}{2}}\left(\eta_{8}-i\eta_{6}\right),\\
\frac{\rmd\alpha_{3}}{\rmd z} = \chi_{2}\alpha_{2}\beta_{2},\\
\frac{\rmd\alpha_{3}^{+}}{\rmd z} = \chi_{2}\alpha_{2}^{+}\beta_{2}^{+},\\
\frac{\rmd\beta_{1}}{\rmd z} = -\chi_{1}\alpha_{1}\alpha_{2},\\
\frac{\rmd\beta_{1}^{+}}{\rmd z} = -\chi_{1}\alpha_{1}^{+}\alpha_{2}^{+},\\
\frac{\rmd\beta_{2}}{\rmd z} = -\chi_{2}\alpha_{2}\alpha_{3}^{+}+\sqrt{\frac{\chi_{2}\alpha_{3}}{2}}\left(\eta_{7}+i\eta_{5}\right),\\
\frac{\rmd\beta_{2}^{+}}{\rmd z} = -\chi_{2}\alpha_{2}^{+}\alpha_{3}+\sqrt{\frac{\chi_{2}\alpha_{3}^{+}}{2}}\left(\eta_{8}+i\eta_{6}\right),
\label{eq:SDEitaly}}
\end{eqnarray}
which we may solve using stochastic integration. The real Gaussian noise terms have the correlations
\begin{equation}
\overline{\eta_{j}(z)}=0,\:\:\:\overline{\eta_{j}(z)\eta_{k}(z')}=\delta_{jk}\delta(z-z').
\end{equation}

\begin{figure}
\begin{center} 
\includegraphics[width=0.8\columnwidth]{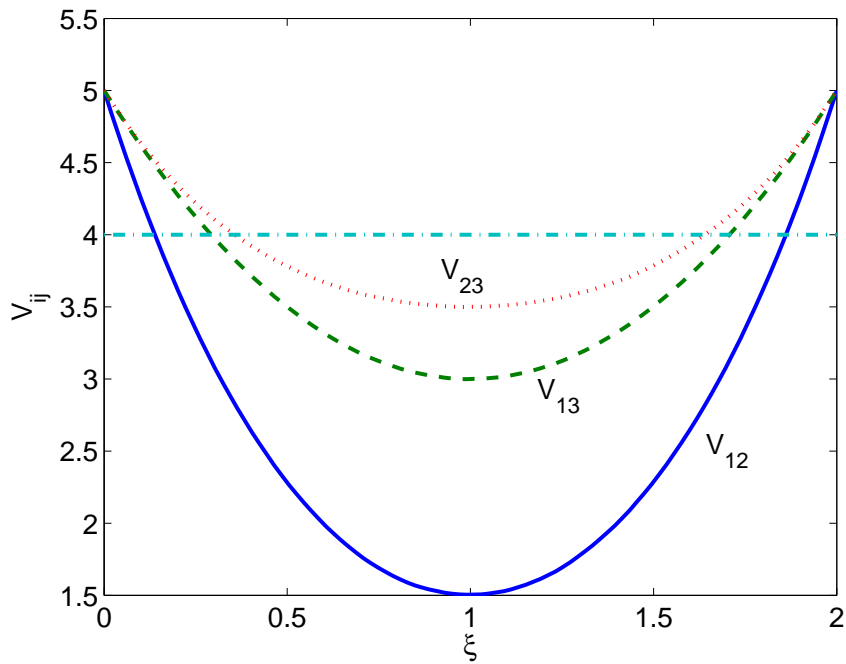}
\end{center} 
\caption{The scheme of reference~\cite{ferraro} with everything symmetric, $\beta_{1}(0)=\beta_{2}(0)=10^{3}$ and $\chi_{1}=\chi_{2}=10^{-2}$, averaged over $1.06\times 10^{6}$ stochastic trajectories. The horizontal axis is the scaled interaction length, $\xi=|\beta_{0}|\chi z$. 
The line at $4$ represents the upper bound for true tripartite entanglement.}
\label{fig:truemilan1}
\end{figure}

\begin{figure}
\begin{center} 
\includegraphics[width=0.8\columnwidth]{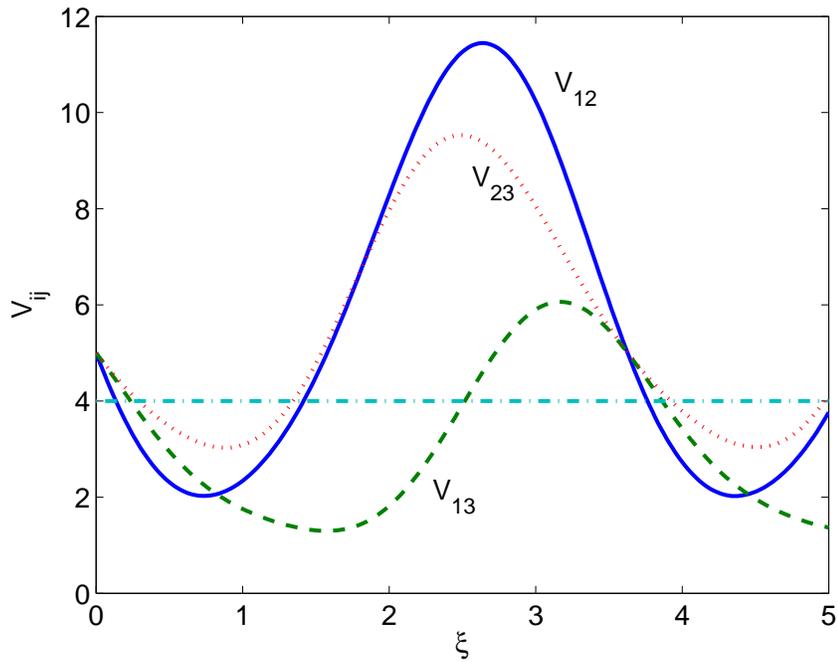}
\end{center} 
\caption{The scheme of reference~\cite{ferraro} as in \fref{fig:truemilan1}, but with $\chi_{2}=2\chi_{1}$, averaged over $3.35\times 10^{6}$ stochastic trajectories.}
\label{fig:truemilan2}
\end{figure}

\begin{figure}
\begin{center} 
\includegraphics[width=0.8\columnwidth]{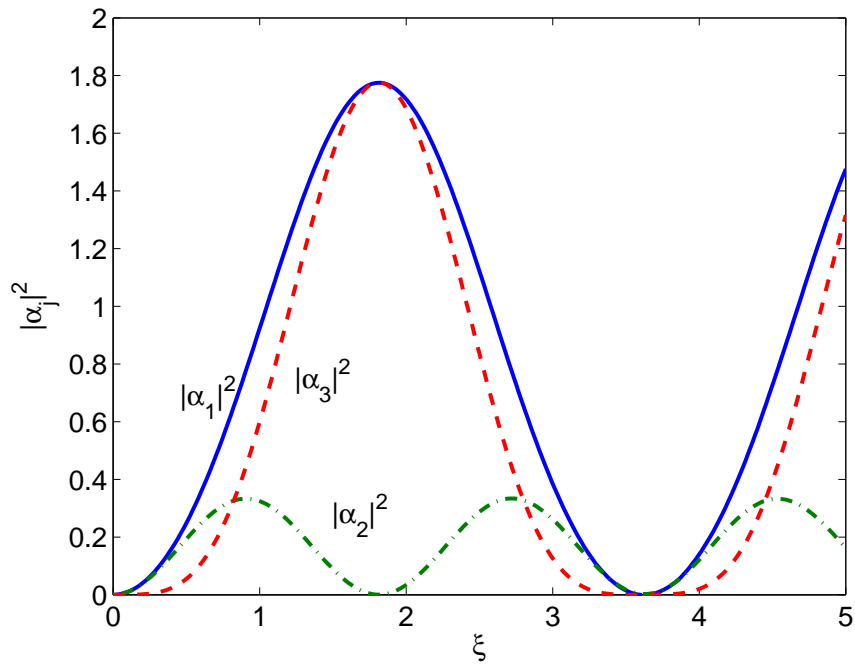}
\end{center} 
\caption{The intensities produced by the interaction of the scheme of reference~\cite{ferraro} for the same parameters as used in \fref{fig:truemilan2}.}
\label{fig:truemilan3}
\end{figure}

The results of stochastic integration in two different parameter regimes are presented in \fref{fig:truemilan1}, \fref{fig:truemilan2} and \fref{fig:truemilan3}. All show genuine tripartite entanglement over some interaction range and all begin with the output modes as vacuum. For the parameters used in \fref{fig:truemilan1}, with $\chi_{1}\beta_{1}(0)=\chi_{2}\beta_{2}(0)$, the output intensities increase monotonically over the range shown and the entanglement disappears. This is  unlike the situation of \fref{fig:truemilan2}, where $\chi_{2}\beta_{2}(0)=2\chi_{1}\beta_{1}(0)$, and both the output fields and the van-Loock Furusawa correlations oscillate over a short interaction length. Although the field intensities for the first situation obviously cannot increase indefinitely, this contrast between monotonically increasing and periodic behaviour of the intensities, depending on the ratios of the pumping and interaction strengths, was mentioned by Smithers and Lu~\cite{Smithers}.

\subsection{Two concurrent nonlinearities}
\label{subsec:besteira}

Another possibility for a travelling wave model is to have a single crystal with two concurrent nonlinearities, each pumped by different modes. This could be achieved either with different polarisations or different frequencies. In this section we will consider a crystal which is pumped at frequencies $\omega_{1}$ and $\omega_{2}$ to produce modes at $\omega_{3},\omega_{4}$ and $\omega_{5}$, where $\omega_{1}=\omega_{3}+\omega_{4}$ and $\omega_{2}=\omega_{4}+\omega_{5}$. 
With all modes quantised with the operators $\hat{b}_{1}(\omega_{1}),\hat{b}_{2}(\omega_{2}),\hat{a}_{1}(\omega_{3}),\hat{a}_{2}(\omega_{4}),\hat{a}_{3}(\omega_{5})$, the interaction Hamiltonian for this scheme becomes
\begin{equation}
H_{\rm int} = \rmi\hbar\left(\chi_{1}\hat{b}_{1}\hat{a}_{1}^{\dag}\hat{a}_{2}^{\dag}+\chi_{2}\hat{b}_{2}\hat{a}_{2}^{\dag}\hat{a}_{3}^{\dag}\right)
+{\rm h.c.},
\label{eq:inventedham} 
\end{equation}
where the $\chi_{j}$ represent the nonlinear interactions. 

We will first examine a simplified analytical model for the propagation of fields described by this Hamiltonian. Assuming perfect phase-matching and ignoring pump depletion, we may set $\gamma_{1}=\chi_{1}\langle \hat{b}_{1}(0)\rangle$ and $\gamma_{2}=\chi_{2}\langle \hat{b}_{2}(0)\rangle$, where the pump fields are initially intense coherent states, to find the Heisenberg equations of motion,
\begin{eqnarray}
\eqalign{
\frac{\rmd\hat{a}_{1}}{\rmd t} = \gamma_{1}\hat{a}_{2}^{\dag},\\
\frac{\rmd\hat{a}_{2}}{\rmd t} = \gamma_{1}\hat{a}_{1}^{\dag}+\gamma_{2}\hat{a}_{3}^{\dag},\\
\frac{\rmd\hat{a}_{3}}{\rmd t} = \gamma_{1}\hat{a}_{2}^{\dag}.
\label{eq:canbesolved}
}
\end{eqnarray}
Setting $\Omega=\sqrt{\gamma_{1}^{2}+\gamma_{2}^{2}}$, we may solve these linear operator equations to find the solutions
\begin{eqnarray}
\fl\eqalign{
\hat{a}_{1}(t) = \hat{a}_{1}(0)\frac{\gamma_{2}^{2}+\gamma_{1}^{2}\cosh\Omega t}{\Omega^{2}}+\hat{a}_{2}^{\dag}(0)\frac{\gamma_{1}\sinh\Omega t}{\Omega}+\hat{a}_{3}(0)\frac{\gamma_{1}\gamma_{2}\left(\cosh\Omega t -1\right)}{\Omega^{2}},\\
\hat{a}_{2}(t) = \hat{a}_{1}^{\dag}(0)\frac{\gamma_{1}\sinh\Omega t}{\Omega}+\hat{a}_{2}(0)\cosh\Omega t+
\hat{a}_{3}^{\dag}(0)\frac{\gamma_{2}\sinh\Omega t}{\Omega},\\
\hat{a}_{3}(t) = \hat{a}_{1}(0)\frac{\gamma_{1}\gamma_{2}\left(\cosh\Omega t-1\right)}{\Omega^{2}}+
\hat{a}_{2}^{\dag}(0)\frac{\gamma_{2}\sinh\Omega t}{\Omega}+\hat{a}_{3}(0)\frac{\gamma_{1}^{2}+\gamma_{2}^{2}\cosh\Omega t}{\Omega^{2}},
\label{eq:analyticresult}
}
\end{eqnarray}
which contain all the information required to calculate the desired correlations. With all the output modes initially vacuum, we find the average intensities
\begin{eqnarray}
\eqalign{
\langle\hat{a}_{1}^{\dag}\hat{a}_{1}\rangle = \frac{\gamma_{1}^{2}\sinh^{2}\Omega t}{\Omega^{2}},\\
\langle\hat{a}_{2}^{\dag}\hat{a}_{2}\rangle = \sinh^{2}\Omega t,\\
\langle\hat{a}_{3}^{\dag}\hat{a}_{3}\rangle = \frac{\gamma_{2}^{2}\sinh^{2}\Omega t}{\Omega^{2}},
\label{eq:analyticN}
}
\end{eqnarray}
with the variances and covariances for the quadratures being found as 
\begin{eqnarray}
\eqalign{
V(\hat{X}_{1}) = V(\hat{Y}_{1}) = 1+2\frac{\gamma_{1}^{2}\sinh^{2}\Omega t}{\Omega^{2}},\\
V(\hat{X}_{2}) = V(\hat{Y}_{2}) = 1+2\sinh^{2}\Omega t,\\
V(\hat{X}_{3}) = V(\hat{Y}_{3}) = 1+2\frac{\gamma_{2}^{2}\sinh^{2}\Omega t}{\Omega^{2}},\\
V(\hat{X}_{1},\hat{X}_{2}) = -V(\hat{Y}_{1},\hat{Y}_{2}) = \frac{\gamma_{1}\sinh 2\Omega t}{\Omega},\\
V(\hat{X}_{1},\hat{X}_{3}) = V(\hat{Y}_{1},\hat{Y}_{3}) = \frac{2\gamma_{1}\gamma_{2}\sinh^{2}\Omega t}{\Omega^{2}},\\
V(\hat{X}_{2},\hat{X}_{3}) = -V(\hat{Y}_{2},\hat{Y}_{3}) = \frac{\gamma_{2}\sinh 2\Omega t}{\Omega}.
\label{eq:misXvars}
}
\end{eqnarray}
We can now calculate the van Loock-Furusawa correlations, finding
\begin{eqnarray}
\fl\eqalign{
V(\hat{X}_{1}-\hat{X}_{2}) = 2\left[1+\left(1+\frac{\gamma_{1}^{2}}{\Omega^{2}}\right)\sinh^{2}\Omega t-\frac{\gamma_{1}}{\Omega}\sinh 2\Omega t\right],\\
V(\hat{X}_{1}-\hat{X}_{3}) = 2\left[1+\left(1-\frac{2\gamma_{1}\gamma_{2}}{\Omega^{2}}\right)\sinh^{2}\Omega t\right],\\
V(\hat{X}_{2}-\hat{X}_{3}) = 2\left[1+\left(1+\frac{\gamma_{2}^{2}}{\Omega^{2}}\right)\sinh^{2}\Omega t-\frac{\gamma_{2}}{\Omega}\sinh 2\Omega t\right],\\
V(\hat{Y}_{1}+\hat{Y}_{2}+\hat{Y}_{3}) = 3+\left(4+\frac{4\gamma_{1}\gamma_{2}}{\Omega^{2}}\right)\sinh^{2}\Omega t - \frac{2(\gamma_{1}+\gamma_{2})}{\Omega}\sinh 2\Omega t,
\label{eq:VLFerro}
}
\end{eqnarray}
which, for $\gamma_{1}=\gamma_{2}=\gamma$, simplify to give
\begin{eqnarray}
\eqalign{
V_{12} = V_{23} = 5+9\sinh^{2}\Omega t-3\sqrt{2}\sinh 2\Omega t,\\
V_{13} = 5+6\sinh^{2}\Omega t-2\sqrt{2}\sinh 2\Omega t,
\label{eq:vLFerro}
}
\end{eqnarray} 
from which we can see that tripartite entanglement is present for shortish interaction times, but does not increase with $\Omega t$. In fact, the minimum of $V_{12}$ and $V_{23}$ is found at $\Omega t = \cosh^{-1}\sqrt{2}$, and has a value of $2$.

We will now turn to stochastic integration of the full equations, without using the undepleted pump approximation.
In what follows we will set the interaction strengths equal, $\chi_{1}=\chi_{2}=\chi$, as these are the conditions which give the maximum of tripartite entanglement for this sytem. 
We define the nonlinear momentum operator for this process as
\begin{equation}
\hat{G}_{nl}(z)= \rmi\hbar\chi\left(
\hat{b}_1^{\dag}\hat{a}_1\hat{a}_2+\hat{b}_2^{\dag}\hat{a}_2\hat{a}_3\right)+{\rm h.c.}.
\label{eq:Gnonlinear2}
\end{equation} 
From this operator we may make a mapping onto a Fokker-Planck equation for the positive-P pseudoprobability distribution, from which we find the It\^o stochastic differential equations,
\begin{eqnarray}
\eqalign{
\frac{\rmd\alpha_{1}}{\rmd z} = \chi\alpha_{2}^{+}\beta_{1}+\sqrt{\frac{\chi\beta_{1}}{2}}\left(\eta_{1}+\rmi\eta_{6}\right),\\
\frac{\rmd\alpha_{1}^{+}}{\rmd z} = \chi\alpha_{2}\beta_{1}^{+}+\sqrt{\frac{\chi\beta_{1}^{+}}{2}}\left(\eta_{2}-\rmi\eta_{5}\right),\\
\frac{\rmd\alpha_{2}}{\rmd z} = \chi\left(\alpha_{1}^{+}\beta_{1}+\alpha_{3}^{+}\beta_{2}\right)+\sqrt{\frac{\chi\beta_{1}}{2}}\left(\eta_{1}-\rmi\eta_{6}\right)+\sqrt{\frac{\chi\beta_{2}}{2}}\left(\eta_{3}+\rmi\eta_{7}\right),\\
\frac{\rmd\alpha_{2}^{+}}{\rmd z} = \chi\left(\alpha_{1}\beta_{1}^{+}+\alpha_{3}\beta_{2}^{+}\right)+\sqrt{\frac{\chi\beta_{1}^{+}}{2}}\left(\eta_{2}+\rmi\eta_{5}\right)+
\sqrt{\frac{\chi\beta_{2}^{+}}{2}}\left(\eta_{4}+\rmi\eta_{8}\right),\\
\frac{\rmd\alpha_{3}}{\rmd z} = \chi\alpha_{2}^{+}\beta_{2}+\sqrt{\frac{\chi\beta_{2}}{2}}\left(\eta_{3}-\rmi\eta_{7}\right),\\
\frac{\rmd\alpha_{3}^{+}}{\rmd z} = \chi\alpha_{2}\beta_{2}^{+}+\sqrt{\frac{\chi\beta_{2}^{+}}{2}}\left(\eta_{4}-\rmi\eta_{8}\right),\\
\frac{\rmd\beta_{1}}{\rmd z} = -\chi\alpha_{1}\alpha_{2},\\
\frac{\rmd\beta_{1}^{+}}{\rmd z} = -\chi\alpha_{1}^{+}\alpha_{2}^{+},\\
\frac{\rmd\beta_{2}}{\rmd z} = -\chi\alpha_{2}\alpha_{3},\\
\frac{\rmd\beta_{2}^{+}}{\rmd z} = -\chi\alpha_{2}^{+}\alpha_{3}^{+},
\label{eq:Itomilan}}
\end{eqnarray}
where the real Gaussian noise terms have the correlations
\begin{equation}
\overline{\eta_{j}(z)}=0,\:\:\overline{\eta_{j}(z)\eta_{k}(z')}=\delta_{jk}\delta(z-z').
\label{eq:notitalians}
\end{equation}
As in the previous section, stochastic averages of products of the variables represent normally ordered operator expectation values, with there being correspondences between $\alpha_{j},\alpha_{j}^{+},\beta_{j},\beta_{j}^{+}$ and $\hat{a}_{j},\hat{a}_{j}^{\dag},\hat{b}_{j},\hat{b}_{j}^{\dag}$.

\begin{figure}
\begin{center} 
\includegraphics[width=0.8\columnwidth]{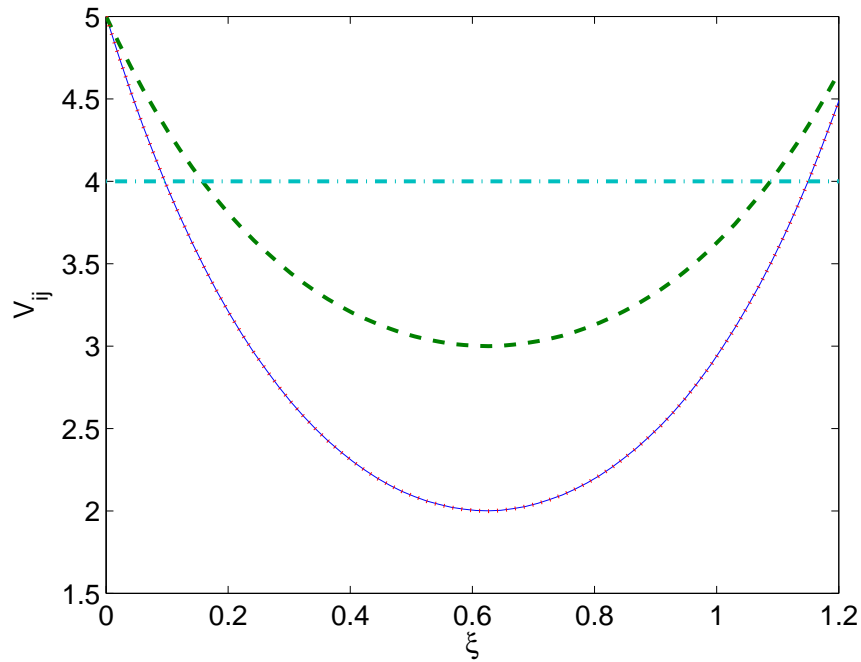}
\end{center} 
\caption{Positive-P solution averaged over $2.68\times 10^6$ stochastic trajectories for the tripartite entanglement criteria with the doubly concurrent Hamiltonian of \sref{subsec:besteira}. The solid line is $V_{12}$ and $V_{23}$, while the dashed line is $V_{13}$. The horizontal axis is the scaled interaction length, $\xi=|\beta_{0}|\chi z$. 
The line at $4$ represents the upper bound for true tripartite entanglement.}
\label{fig:erro}
\end{figure} 

\begin{figure}
\begin{center} 
\includegraphics[width=0.8\columnwidth]{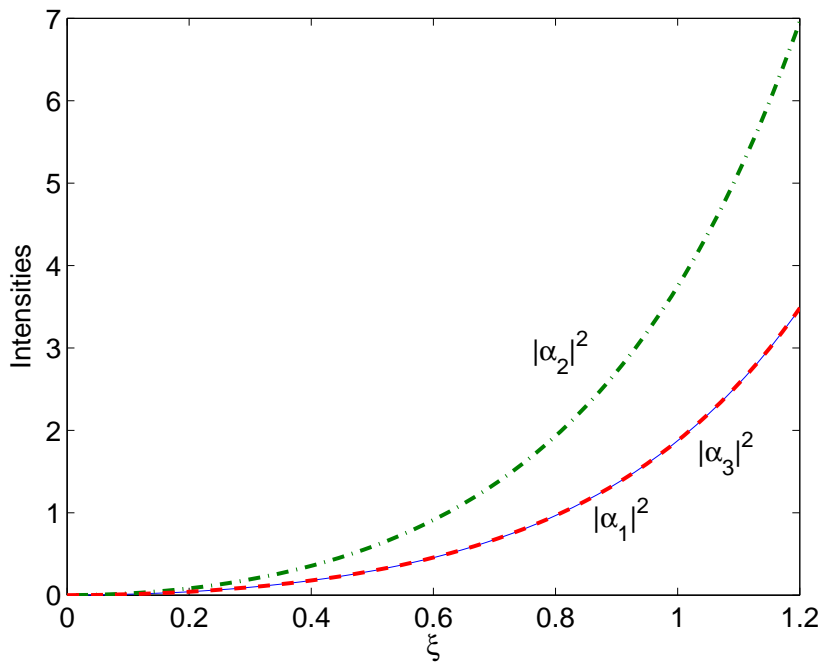}
\end{center} 
\caption{Output intensities for the same system and parameters as \fref{fig:erro}.}
\label{fig:interro}
\end{figure}

The results of stochastic integration of \eref{eq:Itomilan} are shown in \fref{fig:erro} for the tripartite entanglement criteria, for parameter values $\chi=10^{-2}$, $\beta_{1}(0)=\beta_{2}(0)=10^{3}$ and $\alpha_{1}(0)=\alpha_{2}(0)=\alpha_{3}(0)=0$. We see that the correlations are not symmetric, but that tripartite entanglement is available. The output intensities are shown in \fref{fig:interro}, from which we again see that the inequalities are violated for relatively weak fields. For this system, the analytic results which we presented above give the same results as the stochastic integration over the interaction range shown.

\section{Doubled intracavity nonlinearities}
\label{sec:cavity2}

Now that we have demonstrated that the two Hamiltonians of \sref{sec:quitegood} can produce entanglement, we will examine the more realistic physical situations where the processes described happen inside pumped optical cavities. As optical cavities can be tuned to be resonant with only a certain number of modes, a fuller description of the physics involved can be given in a simple manner. 

\subsection{Cascaded nonlinearities in a cavity}
\label{subsec:cavFerraro}

To consider the scheme described by the interaction Hamiltonian of \eref{eq:parma} inside a resonant pumped optical cavity, we must add terms to the Hamiltonian, so that
\begin{equation}
H=H_{\rm int}+H_{\rm pump}+H_{\rm damp},
\label{eq:cavmilano}
\end{equation}
where $H_{\rm int}$ has the same form as the interaction Hamiltonian of \eref{eq:parma}, and
\begin{eqnarray}
H_{\rm pump} &=& \rmi\hbar\left(\epsilon_{1}\hat{b}_{1}^{\dag}+\epsilon_{2}\hat{b}_{2}^{\dag}\right) + {\rm h.c.}\nonumber\\
H_{\rm damp} &=& \hbar\left(\sum_{k}\hat{\Gamma}_{a}^{k}\hat{a}_{k}^{\dag}+\sum_{j}\hat{\Gamma}_{b}^{j}\hat{b}_{j}^{\dag}\right) + {\rm h.c},
\label{eq:cavparma}
\end{eqnarray}
where $k=1,2,3$ and $j=1,2$. In the above, the $\epsilon_{j}$ are the classical pump amplitudes at the two input frequencies and the $\Gamma_{a,b}^{k}$ are bath operators. The field operators now refer to the intracavity fields. This now becomes equivalent to the scheme of Guo \etal~\cite{Guo}, who analysed it using quantum Langevin equations in an undepleted pumps approximation. Following the same procedures used in deriving \eref{eq:SDEitaly} and making the usual zero-temperature Markovian bath approximation, we find the positive-P equations which give a full quantum description of this system, 
\begin{eqnarray}
\eqalign{
\frac{\rmd\alpha_{1}}{\rmd t} = -\gamma_{1}\alpha_{1} + \chi_{1}\alpha_{2}^{+}\beta_{1}+\sqrt{\frac{\chi_{1}\beta_{1}}{2}}\left(\eta_{1}+i\eta_{3}\right),\\
\frac{\rmd\alpha_{1}^{+}}{\rmd t} = -\gamma_{1}\alpha_{1}^{+} + \chi_{1}\alpha_{2}\beta_{1}^{+}+\sqrt{\frac{\chi_{1}\beta_{1}^{+}}{2}}\left(\eta_{2}+i\eta_{4}\right),\\
\frac{\rmd\alpha_{2}}{\rmd t} = -\gamma_{2}\alpha_{2} + \chi_{1}\alpha_{1}^{+}\beta_{1}-\chi_{2}\alpha_{3}\beta_{2}^{+}+\sqrt{\frac{\chi_{1}\beta_{1}}{2}}\left(\eta_{1}-i\eta_{3}\right)-
\sqrt{\frac{\chi_{2}\alpha_{3}}{2}}\left(\eta_{7}-i\eta_{5}\right),\\
\frac{\rmd\alpha_{2}^{+}}{\rmd t} = -\gamma_{2}\alpha_{2}^{+}+ \chi_{1}\alpha_{1}\beta_{1}^{+}-\chi_{2}\alpha_{3}^{+}\beta_{2}+\sqrt{\frac{\chi_{1}\beta_{1}^{+}}{2}}\left(\eta_{2}-i\eta_{4}\right)-
\sqrt{\frac{\chi_{2}\alpha_{3}^{+}}{2}}\left(\eta_{8}-i\eta_{6}\right),\\
\frac{\rmd\alpha_{3}}{\rmd t} = -\gamma_{3}\alpha_{3} + \chi_{2}\alpha_{2}\beta_{2},\\
\frac{\rmd\alpha_{3}^{+}}{\rmd t} = -\gamma_{3}\alpha_{3}^{+} +\chi_{2}\alpha_{2}^{+}\beta_{2}^{+},\\
\frac{\rmd\beta_{1}}{\rmd t} = \epsilon_{1}-\kappa_{1}\beta_{1}-\chi_{1}\alpha_{1}\alpha_{2},\\
\frac{\rmd\beta_{1}^{+}}{\rmd t} = \epsilon_{1}^{\ast}-\kappa_{1}\beta_{1}^{+}-\chi_{1}\alpha_{1}^{+}\alpha_{2}^{+},\\
\frac{\rmd\beta_{2}}{\rmd t} = \epsilon_{2}-\kappa_{2}\beta_{2}-\chi_{2}\alpha_{2}\alpha_{3}^{+}+\sqrt{\frac{\chi_{2}\alpha_{3}}{2}}\left(\eta_{7}+i\eta_{5}\right),\\
\frac{\rmd\beta_{2}^{+}}{\rmd t} = \epsilon_{2}^{\ast}-\kappa_{2}\beta_{2}^{+}-\chi_{2}\alpha_{2}^{+}\alpha_{3}+\sqrt{\frac{\chi_{2}\alpha_{3}^{+}}{2}}\left(\eta_{8}+i\eta_{6}\right).
\label{eq:SDEcavitaly}}
\end{eqnarray}
Note that the correlations of the noise terms are now in time rather than in the spatial variable.
These equations can be integrated numerically in any parameter regime, including near to any critical points of the system.
However, as we will not be concerned with the behaviour of this system in the neighbourhood of any critical points, we will proceed via a linearised fluctuation analysis. This involves separating the variables of the positive-P equations into their mean-field steady-state solutions plus a fluctuating part, e.g. $\alpha_{1}=\alpha_{1}^{ss}+\delta\alpha_{1}$. Solving the classical equations of motion to find the steady-state solutions, we may then write equations for the fluctuations from which we can calculate the output spectral quantities of interest~\cite{Danbook}. 
Neglecting the noise terms in \eref{eq:SDEcavitaly}, we find the following classical equations for the interacting fields,
\begin{eqnarray}
\frac{\rmd\alpha_{1}}{\rmd t} &=& -\gamma_{1}\alpha_{1} + \chi_{1}\alpha_{2}^{\ast}\beta_{1},\nonumber\\
\frac{\rmd\alpha_{2}}{\rmd t} &=& -\gamma_{2}\alpha_{2} + \chi_{1}\alpha_{1}^{\ast}\beta_{1}-\chi_{2}\alpha_{3}\beta_{2}^{\ast},\nonumber\\
\frac{\rmd\alpha_{3}}{\rmd t} &=& -\gamma_{3}\alpha_{3} + \chi_{2}\alpha_{2}\beta_{2},\nonumber\\
\frac{\rmd\beta_{1}}{\rmd t} &=& \epsilon_{1}-\kappa_{1}\beta_{1}-\chi_{1}\alpha_{1}\alpha_{2},\nonumber\\
\frac{\rmd\beta_{2}}{\rmd t} &=& \epsilon_{2}-\kappa_{2}\beta_{2}-\chi_{2}\alpha_{2}\alpha_{3}^{\ast},
\label{eq:cavmilanoclass}
\end{eqnarray}
where the $\gamma_{j}\;(\kappa_{j})$ are the cavity damping rates for the $\alpha_{j}\;(\beta_{j})$.

We see that one possible set of solutions to these equations is
\begin{eqnarray}
\alpha_{j}^{ss} = 0,\\
\beta_{j}^{ss} = \epsilon_{j}/\kappa_{j},
\label{eq:cavidademilano}
\end{eqnarray}
which are reminiscent of those found for the well-known OPO below threshold. To examine the stability of these solutions, we write the linearised equation for the fluctuations,
\begin{equation}
\rmd\,\delta\tilde{\alpha} = A\delta\tilde{\alpha}\,\rmd t+B\,\rmd W,
\label{eq:OUequation}
\end{equation}
where $\tilde{\alpha}=\left[\delta\alpha_{1},\delta\alpha_{1}^{+},\delta\alpha_{2},\delta\alpha_{2}^{+},\delta\alpha_{3},\delta\alpha_{3}^{+},\delta\beta_{1},
\delta\beta_{1}^{+},\delta\beta_{2},
\delta\beta_{2}^{+}\right]^{T}$, $B$ is the matrix of the noise terms of \eref{eq:SDEitaly}, but with the classical steady-state solutions used in place of the stochastic variables and $\rmd W$ is a vector of Wiener increments. The drift matrix is found as
\begin{equation}
A = \left[A_{1}\: A_{2}\right],
\end{equation}
where
\begin{equation}
\fl 
A_{1}=\left[\begin{array}{cccccc}
-\gamma_{1} & 0 & 0 & \chi_{1}\beta_{1}^{ss} & 0 & 0 \\
0 & -\gamma_{1} & \chi_{1}(\beta_{1}^{\ast})^{ss} & 0 & 0 & 0 \\
0 & \chi_{1}\beta_{1}^{ss} & -\gamma_{2} & 0 & -\chi_{2}(\beta_{2}^{\ast})^{ss} & 0 \\
\chi_{1}(\beta_{1}^{\ast})^{ss} & 0  & 0 & -\gamma_{2} & 0 & -\chi_{2}\beta_{2}^{ss} \\
0 & 0 & \chi_{2}\beta_{2}^{ss} & 0 & -\gamma_{3} & 0 \\
0 & 0 & 0 & \chi_{2}(\beta_{2}^{\ast})^{ss} & 0 & -\gamma_{3}\\
-\chi_{1}\alpha_{2}^{ss} & 0 & -\chi_{1}\alpha_{1} & 0 & 0 & 0\\
0 & -\chi_{1}(\alpha_{2}^{\ast})^{ss} & 0 & -\chi_{1}(\alpha_{1}^{\ast})^{ss} & 0 & 0 \\
0 & 0 & -\chi_{2}(\alpha_{3}^{\ast})^{ss} & 0 & 0 & -\chi_{2}\alpha_{2}^{ss} \\
0 & 0 & 0 & -\chi_{2}\alpha_{3}^{ss} & -\chi_{2}(\alpha_{2}^{\ast})^{ss} & 0
\end{array}\right],
\label{eq:Amatmilan1}
\end{equation}
and
\begin{equation} 
A_{2}=\left[\begin{array}{cccc}
\chi_{1}(\alpha_{2}^{\ast})^{ss} & 0 & 0 & 0 \\
0 & \chi_{1}\alpha_{2}^{ss} & 0 & 0 \\
\chi_{1}(\alpha_{1}^{\ast})^{ss} & 0 & 0 & -\chi_{2}\alpha_{3}^{ss} \\
0 & \chi_{1}\alpha_{1}^{ss} & -\chi_{2}(\alpha_{3}^{\ast})^{ss} & 0 \\
\chi_{2}\alpha_{2}^{ss} & 0 & 0 & 0 \\
0 & \chi_{2}(\alpha_{2}^{\ast})^{ss} & 0 & 0 \\
-\kappa_{1} & 0 & 0 & 0 \\
0 & -\kappa_{1} & 0 & 0 \\
0 & 0 & -\kappa_{2} & 0 \\
0 & 0 & 0 & -\kappa_{2}
\end{array}\right].
\label{eq:Amatmilan2}
\end{equation}
As long as none of the eigenvalues of this drift matrix have a positive real part, the solutions are stable and the linearised fluctuation analysis should be valid. Although general analytical expressions for the eigenvalues can be found, these are rather complicated. In the simplifying case below threshold where we set the output loss rates equal, $\gamma_{j}=\gamma$ and $\kappa_{j}=\kappa$, we find a degenerate eigenvalue which can be positive,
\begin{equation}
\lambda = -\gamma+\frac{\sqrt{\chi_{1}^{2}\epsilon_{1}^{2}-\chi_{2}^{2}\epsilon_{2}^{2}}}{\kappa}.
\label{eq:italianeigenvalue}
\end{equation}
This eigenvalue sets a condition for the relative strengths of the pumps,
\begin{equation}
\chi_{1}^{2}\epsilon_{1}^{2}-\chi_{2}^{2}\epsilon_{2}^{2}<\gamma^{2}\kappa^{2}.
\label{eq:algonovo}
\end{equation}
When this condition is violated, the below threshold solutions are unstable.
 
\begin{figure}
\begin{center}\includegraphics[width=0.8\columnwidth]{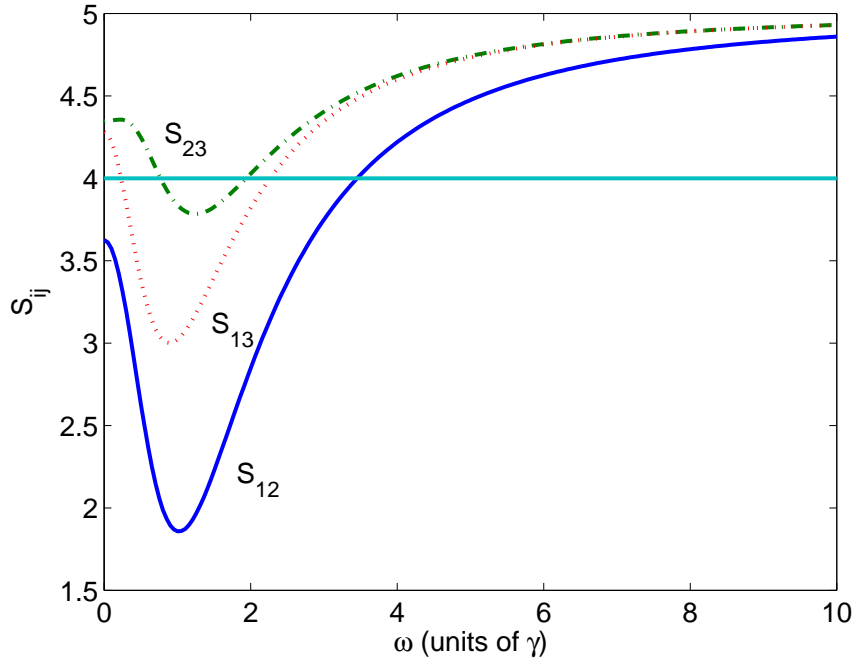}
\end{center}
\caption{Tripartite entanglement criteria for the system of section~\ref{subsec:cavFerraro}, for $\kappa=\gamma=1$, $\chi_{1}=\chi_{2}=10^{-2}$ and $\epsilon_{1}=\epsilon_{2}=0.9\gamma\kappa/\chi_{1}$.}
\label{fig:cavferraro}
\end{figure}

The spectral correlations are found in the normal manner, via the equation
\begin{equation}
S(\omega)=\left(A+i\omega\right)^{-1}D\left(A^{T}-i\omega\right)^{-1},
\label{eq:OUstandard}
\end{equation}
where $D$ is the diffusion matrix with the steady-state values of the fields,
and the standard input-output relationships~\cite{inout}. 
We will denote the output spectral correlations equivalent to the $V_{ij}$ of \eref{eq:tripart} as $S_{ij}$.
Although we were able to find analytic expressions for the correlations, these were exremely complicated and gave little insight, therefore we have presented numerical results in \fref{fig:cavferraro}. Our numerical investigations over a range of parameters did not find any tripartite entanglement noticeably better than that presented. We found that when the pumping rates increase above the value of $\epsilon_{j}=\gamma\kappa/\chi_{j}$, the entanglement rapidly disappears.
What is immediately visible is that the inequalities are not all violated equally, although true tripartite entanglement is demonstrated. We note that numerical investigations show that this system exhibits a range of behaviours, with self-pulsing type oscillations and possible bistability for particular parameter regimes, but that here we are only interested in its suitability as a source of tripartite entanglement and an investigation of these effects is outside the scope of this work.

\subsection{Intra-cavity concurrent twin nonlinearities}
\label{subsec:caverro}

In a similar manner to the preceding section (\ref{subsec:cavFerraro}), we will now investigate the performance of an intracavity version of the system described by 
\eref {eq:inventedham}. We again find the classical equations by dropping the noise terms in the appropriate positive-P equations and solve these for the mean fields in the steady-state. The equations for the mean fields are
\begin{eqnarray}
\frac{\rmd\alpha_{1}}{\rmd t} &=& -\gamma_{a}\alpha_{1}+\chi\alpha_{2}^{\ast}\beta_{1},\nonumber\\
\frac{\rmd\alpha_{2}}{\rmd t} &=& -\gamma_{a}\alpha_{2}+\chi\left(\alpha_{1}^{\ast}\beta_{1}+\alpha_{3}^{\ast}\beta_{2}\right),\nonumber\\
\frac{\rmd\alpha_{3}}{\rmd t} &=& -\gamma_{a}\alpha_{3}+\chi\alpha_{2}^{\ast}\beta_{2},\nonumber\\
\frac{\rmd\beta_{1}}{\rmd t} &=& \epsilon_{1}-\gamma_{b}\beta_{1}-\chi\alpha_{1}\alpha_{2},\nonumber\\
\frac{\rmd\beta_{2}}{\rmd t} &=& \epsilon_{2}-\gamma_{b}\beta_{2}-\chi\alpha_{2}\alpha_{3},
\label{eq:sscaverro}
\end{eqnarray}
where the $\gamma_{a,b}$ are the cavity damping rates for the appropriate modes and the $\epsilon_{j}$ are the classical pumping terms. Although we have chosen the simple case where both pump modes have the same loss rate, as do the three signal modes, this is not essential, although it does serve to simplify our analysis. We may now solve the above equations to find the steady-state mean field solutions and the conditions for the stability of the linearised fluctuation analysis.

To find analytical expressions, we will set $\epsilon_{1}=\epsilon_{2}=\epsilon$. We find that there is an oscillation threshold at the value $\epsilon_{\rm th}=\gamma_{a}\gamma_{b}/2\chi$, below which $\alpha_{1}^{ss}=\alpha_{2}^{ss}=\alpha_{3}^{ss}=0$ and $\beta_{1}^{ss}=\beta_{2}^{ss}=\epsilon/\gamma_{b}$. Above this threshold, we find
\begin{eqnarray}
\alpha_{2}^{ss} &=& \pm\sqrt{\frac{2}{\chi}(\epsilon-\epsilon_{\rm th})},\nonumber\\
\alpha_{1}^{ss} &=& \alpha_{3}^{ss} = \pm\sqrt{\frac{1}{\chi}(\epsilon-\epsilon_{\rm th})},\nonumber\\
\beta_{1}^{ss} &=& \beta_{2}^{ss} = \frac{\gamma_{a}}{2\chi},
\label{eq:ssmistake} 
\end{eqnarray}
where all the $\alpha_{j}$ must have the same sign. What is unusual about this system in comparison with the normal optical parametric oscillator is that it is stable at threshold, with the critical point for the below threshold solutions being at a pump amplitude of $\epsilon_{c}=\gamma_{a}\gamma_{b}/\sqrt{2}\chi$, which means that $\epsilon_{c}=\sqrt{2}\epsilon_{\rm th}$. For this pumping strength, the actual above threshold solutions are stable, so that this system, at least for the parameter regimes we consider here, is always stable.  
With these steady-state solutions, 
we have all the information we need to calculate the output spectral correlations of interest.

\begin{figure}
\begin{center}\includegraphics[width=0.8\columnwidth]{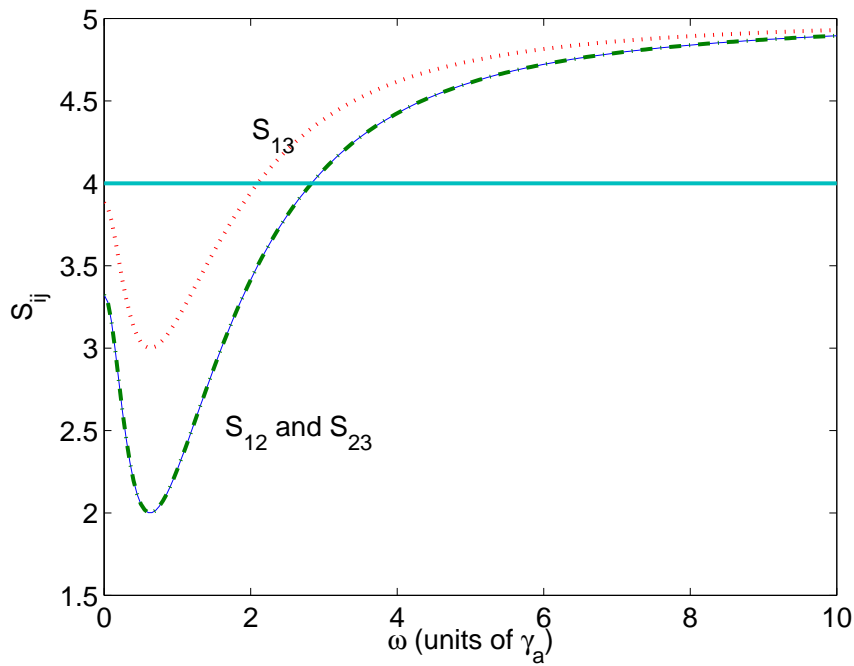}
\end{center}
\caption{Tripartite entanglement criteria for the system of section~\ref{subsec:caverro}, for $\gamma_{a}=\gamma_{b}=1$, $\chi=10^{-2}$ and $\epsilon=0.9\epsilon_{\rm th}$.}
\label{fig:cavmistake}
\end{figure}

\begin{figure}
\begin{center}\includegraphics[width=0.8\columnwidth]{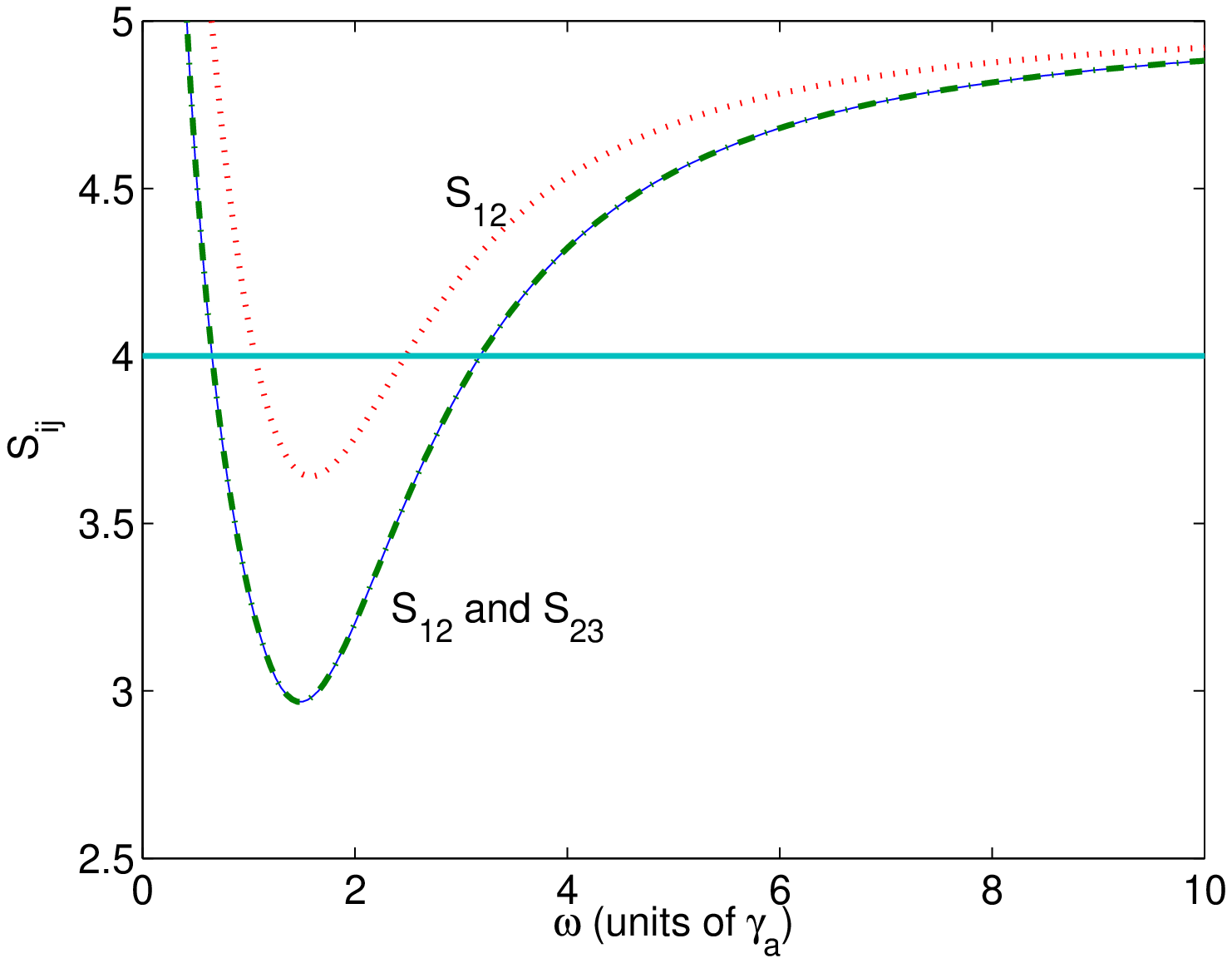}
\end{center}
\caption{Tripartite entanglement criteria for the system of section~\ref{subsec:caverro}, for $\gamma_{a}=\gamma_{b}=1$, $\chi=10^{-2}$ and $\epsilon=2\epsilon_{\rm th}$.}
\label{fig:cavacima}
\end{figure}

Below the oscillation threshold we can find relatively simple expressions for the output spectral correlations,
\begin{eqnarray}
\fl\eqalign{
S_{12}(\omega) = S_{23}(\omega) = 5 - \frac{24\gamma_{a}\gamma_{b}\chi\epsilon\left(\gamma_{a}^{2}\gamma_{b}^{2}-3\gamma_{a}\gamma_{b}\chi\epsilon+2\chi^{2}\epsilon^{2}+
\gamma_{b}^{2}\omega^{2}\right)}{\gamma_{a}^{4}\gamma_{b}^{4}+\left(2\chi^{2}\epsilon^{2}+\gamma_{b}^{2}\omega^{2}\right)^{2}+
2\gamma_{a}^{2}\gamma_{b}^{2}\left(\gamma_{b}^{2}\omega^{2}-2\chi^{2}\epsilon^{2}\right)},\\
S_{13}(\omega) = 5 - \frac{16\gamma_{a}\gamma_{b}\chi\epsilon\left(\gamma_{a}^{2}\gamma_{b}^{2}-3\gamma_{a}\gamma_{b}\chi\epsilon+2\chi^{2}\epsilon^{2}+
\gamma_{b}^{2}\omega^{2}\right)}{\gamma_{a}^{4}\gamma_{b}^{4}+\left(2\chi^{2}\epsilon^{2}+\gamma_{b}^{2}\omega^{2}\right)^{2}+
2\gamma_{a}^{2}\gamma_{b}^{2}\left(\gamma_{b}^{2}\omega^{2}-2\chi^{2}\epsilon^{2}\right)},
\label{eq:mistakebelow}
}
\end{eqnarray}
whereas above threshold the analytical expressions become more complicated.

We show results for the tripartite criteria in \fref{fig:cavmistake} immediately below the oscillation threshold, and in \fref{fig:cavacima} for $\epsilon=2\epsilon_{\rm th}$. We find that the tripartite entanglement is strongest at threshold, with $S_{12}=S_{23}$ and $S_{13}$ showing a lesser violation of the criterion. We also found that, well above threshold, the individual quadratures $X_{1}$ and $X_{3}$ become squeezed, although $X_{2}$ remains above the shot-noise level.
  

\section{Conclusions}

We have examined two different interaction schemes in terms of their potential for creating continuous variable tripartite entanglement. One of these is based on two cascaded nonlinearities and the other on two concurrent nonlinearities. Although the scheme based on cascaded linearities has been analysed previously, we have extended these analyses to include depletion of the pumping fields. Both these schemes were shown to exhibit approximately the same degree of tripartite entanglement, and are therefore suitable candidates for practical applications. The concurrent scheme has the possible advantage that two of the outputs are of equal intensity and is symmetric in two of the correlations, whereas the cascaded scheme produces three different intensities and all three correlations give different values. Which one is preferable in a given situation may come down to the preferences and experiences of the experimenters wishing to use them. 

\ack

This research was supported by the Australian Research Council and the Queensland State Government.

  

\Bibliography{99}

\bibitem{Jing}{Jing J, Zhang J, Yan Y, Zhao F,  Xie C and  Peng K 2003 \PRL {\bf 90} 167903}
%
\bibitem{aoki}{Aoki T, Takei N, Yonezawa H, Wakui K, Hiraoka T and Furusawa A 2003 \PRL {\bf 91} 080404}
%
\bibitem{Guo}{Guo J, Zou H, Zhai Z, Zhang J and Gao J 2005 \PRA {\bf 71} 034305}
%
\bibitem{ferraro}{Ferraro A, Paris M G A, Bondani M, Allevi A, Puddu E and Andreoni A 2004 \JOSAB {\bf 21} 1241}
%
\bibitem{Nosso}{Bradley A S, Olsen M K, Pfister O and Pooser R C 2005 \PRA in press}
%
\bibitem{vanLoock2003}{van Loock P and Furusawa A 2003 \PRA {\bf 67} 052315}
%
\bibitem{Duan}{Duan L M, Giedke G, Cirac J I and Zoller P 2000 \PRL {\bf 84} 2722}
%
\bibitem{GHZ1}{Greenberger D M, Horne M A and Zeilinger A \emph{in Bell's Theorem, Quantum Theory, and Conceptions of the Universe, ed. M. Kafatos} (Kluwer Academic, Dordrecht, 1989)}
%
\bibitem{GHZ2}{Greenberger D M, Horne M A, Shimony A and Zeilinger A 1990 {\it Am. J. Phys.} {\bf 58} 1131}
%
\bibitem{fedorente}{Drummond P D 1990 \PRA {\bf 42} 6845}
%
\bibitem{Smithers}{Smithers M E and Lu E Y C 1974 \PRA {\bf 10} 1874}
%
\bibitem{frenchconnection}{Ferraro A and Paris M G A 2005 \JOB {\bf 7} 174}
%
\bibitem{revive}{Olsen M K, Horowicz R J, Plimak L I, Treps N and C. Fabre C 2000 \PRA {\bf 61} 021803}
%
\bibitem{turco}{Olsen M K, Plimak L I and Khoury A Z 2003 \OC {\bf 215} 101}
%
\bibitem{shgepr}{Olsen M K 2004 \PRA {\bf 70} 035801}
%
\bibitem{Huttner}{Huttner B, Serulnik S and Ben-Aryeh Y 1990 \PRA {\bf 42} 5594}
%
\bibitem{Carlton}{Caves C M and Crouch D D 1987 \JOSAB {\bf 4} 1535}
%
\bibitem{P+}{Drummond P D and Gardiner C W 1980 \JPA {\bf 13} 2353}
%
\bibitem{Shen}{Shen Y R 1967 \PR {\bf 155} 921}
%
\bibitem{Crispin}{Gardiner C W \emph{Quantum Noise} (Springer, Berlin,
1991)}
%
\bibitem{Danbook}{Walls D F and Milburn G J \emph{Quantum Optics} (Springer-Verlag, Berlin, 1994)}
%
\bibitem{inout}{Gardiner C W and Collett M J 1985 \PRA {\bf 31} 3761.}
%

\endbib
\end{document}